\documentclass[aps,prd,nofootinbib,amsfonts,notitlepage,longbibliography]{revtex4-1}
\usepackage[utf8]{inputenc}
\usepackage[T1]{fontenc}
\usepackage{hyperref}
\usepackage{graphicx,bm,amsmath,color}
\usepackage{bbold}
\usepackage{wasysym}
\usepackage{verbatim}

\newcommand{\be}{\begin{equation}}
\newcommand{\ee}{\end{equation}}
\newcommand{\beq}{\begin{eqnarray}}
\newcommand{\eeq}{\end{eqnarray}}
\newcommand{\ba}{\begin{align}}
\newcommand{\ea}{\end{align}}

\newcommand{\Utgu}{U_{\theta_G^{(1)}}}
\newcommand{\Utg}{U_{\theta_G}}
\newcommand{\Tgu}{\theta_G^{(1)}}
\newcommand{\Tg}{\theta_G}

\newcommand{\cgbb}{TLMBB}

\begin{document}

\title{Before spacetime: A proposal of a framework for multiverse quantum cosmology based on three cosmological conjectures}
\author{J.L. Alonso}
\affiliation{Departamento de F\'{\i}sica Te\'orica,
Universidad de Zaragoza, Zaragoza 50009, Spain}
\author{J.M. Carmona}
\affiliation{Departamento de F\'{\i}sica Te\'orica,
Universidad de Zaragoza, Zaragoza 50009, Spain}

\begin{abstract}
The three cosmological conjectures to which our work refers are: the phenomenon called geodesic incompleteness, the physical gravitational $\theta_G$-term that would characterize the $1$-parameter family of inequivalent vacua of quantum gravidynamics, and the hypothesis of multiversality~\cite{Wilczek:2013lra}, more specifically, a zero-energy multiverse.

The known cosmological phenomenology leads under plausible assumptions to theorems which establish that the universe is past incomplete~\cite{Borde:2001nh,Vilenkin:2014yva}. Here, starting from Wilczek's definition of multiverse~\cite{Wilczek:2013lra} (a larger physical structure of which the universe forms part) and that spacetime is much larger than the observable universe, in a new sense suggested by these theorems, we place the observable universe, labelled by $\Utgu$, within a multiverse ensemble, \{$\Utg$\}. Its topological $\Tgu$-term would characterize the observable universe from the Planck epoch until the present time, and it could have physical effects in, for example, black-hole physics.

Our proposal is therefore a possible framework for a multiverse quantum cosmology, in which the temporal parameters (see figures in the main text) start from a ``timeless multiverse big bang'' (TLMBB), where all members of the multiverse ensemble, \{$\Utg$\}, disappear, together with their corresponding classical spacetimes. Since quantum cosmology can be viewed as one attempt among many to face with the question of finding a gravitational quantum theory, if the TLMBB were the appropriate ground to define the physical or mathematical underlying structure of quantum cosmology, then multiversality could come to have a predictive power within our observable universe.
\end{abstract}

\maketitle

\section{Objective and introduction}
\label{sec:intro}

It is not unusual that plausible conjectures appear in the course of the development of a scientific discipline. At some point, framing them in a common framework (if one hits it right) can help to glimpse other pieces of the ``puzzle'', making possible to see whether some proposed theoretical approaches fit well in that framework. This could be the case, for example, of a gravitational quantum theory (QG) .

Following this spirit, in this letter we try to place in a common framework three plausible conjectures that appear to be more plausible all together than separate: the phenomenon called geodesic incompleteness~\cite{Borde:2001nh,Vilenkin:2014yva}, the 
conjectured physical gravitational topological $\Tg$-term that would characterize the $1$-parameter family of inequivalent vacua of quantum gravidynamics (QGD)~\cite{Deser:1980kc,Ashtekar:1988sw,Calcagni:2017sdq}, and the hypothesis of multiversality~\cite{Wilczek:2013lra}, more specifically, a zero-energy multiverse (see Sect.~\ref{sec:zero}). This framework would then constitute a natural proposal for a quantum cosmology (QC) based on the idea of multiversality, that is, for a multiverse quantum cosmology (MQC).

Our boot conjecture is based on the past-incompleteness theorems whose hypotheses are clearly specified in Refs.~\cite{Borde:2001nh,Vilenkin:2014yva,Calcagni:2017sdq}. Roughly speaking, these theorems establish the need of a singularity in spacetime at the classical level for expanding universes. It is in the context of these theorems, valid in classical spacetime, that we speak of the big bang singularity. As Stephen Hawking explains, referring to open questions on the origin of the universe on page 85 of Ref.~\cite{Hawking:2005yn}: ``The general theory of relativity, on its own, cannot explain these features or answer these questions. This is because it predicts that the universe started off with infinite density at the big bang singularity. At the singularity, general relativity and all other physical laws would break down. One cannot predict what would come out of the singularity. As I explained before, this means that one might as well cut any events before the big bang out of the theory, because they can have no effect on what we observe. Spacetime would have a boundary---a beginning at the big bang''.
Having said that, it is necessary to keep in mind that singularity avoidance can be obtained in models of QC (see, e.g., Ref.~\cite{Albarran:2016ewi} and references therein).

Now we put the focus on multiversality and we make concrete our conjecture: we will speak of a timeless multiverse big bang (TLMBB), whose meaning is precisely to cut any events before the big bang out of the theory (so that spacetime would have a boundary---a beginning at the big bang---), but which produces a much larger spacetime, \{$\Utg$\}, than the observable universe, in a new sense suggested by the incompleteness theorems~\cite{Borde:2001nh,Vilenkin:2014yva}.

\begin{figure}%
\includegraphics[scale=0.5]{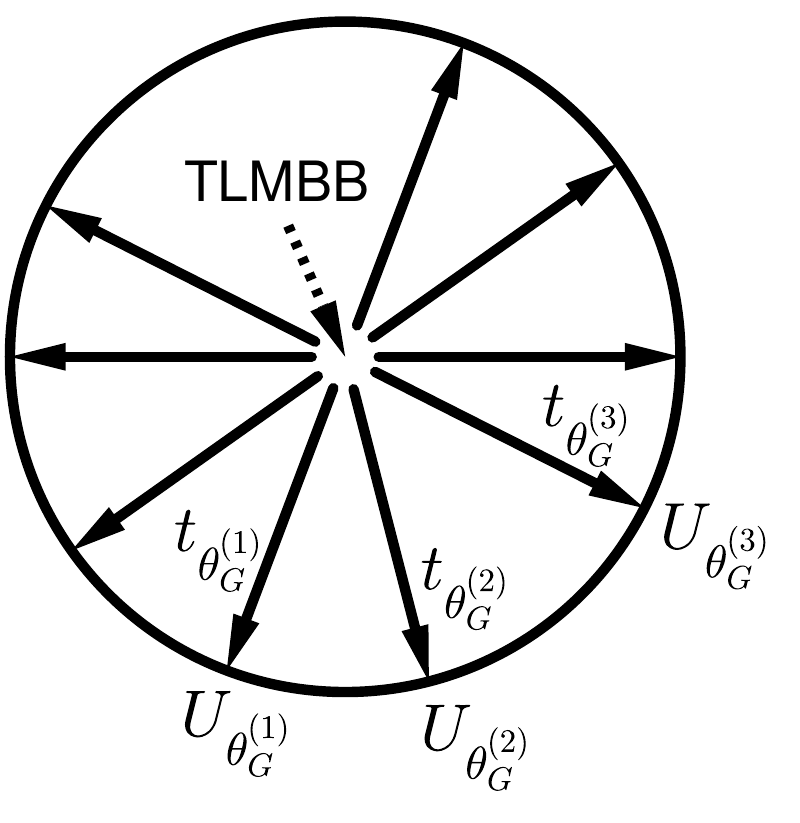}%
\caption{Plot of the temporal parameters, $t_{\theta_G^{(i)}}$, of the multiverse ensemble. Each one corresponds to a universe $U_{\theta_G^{(i)}}$ (see text).}%
\label{fig1}%
\end{figure}

\begin{figure}%
\includegraphics[scale=0.5]{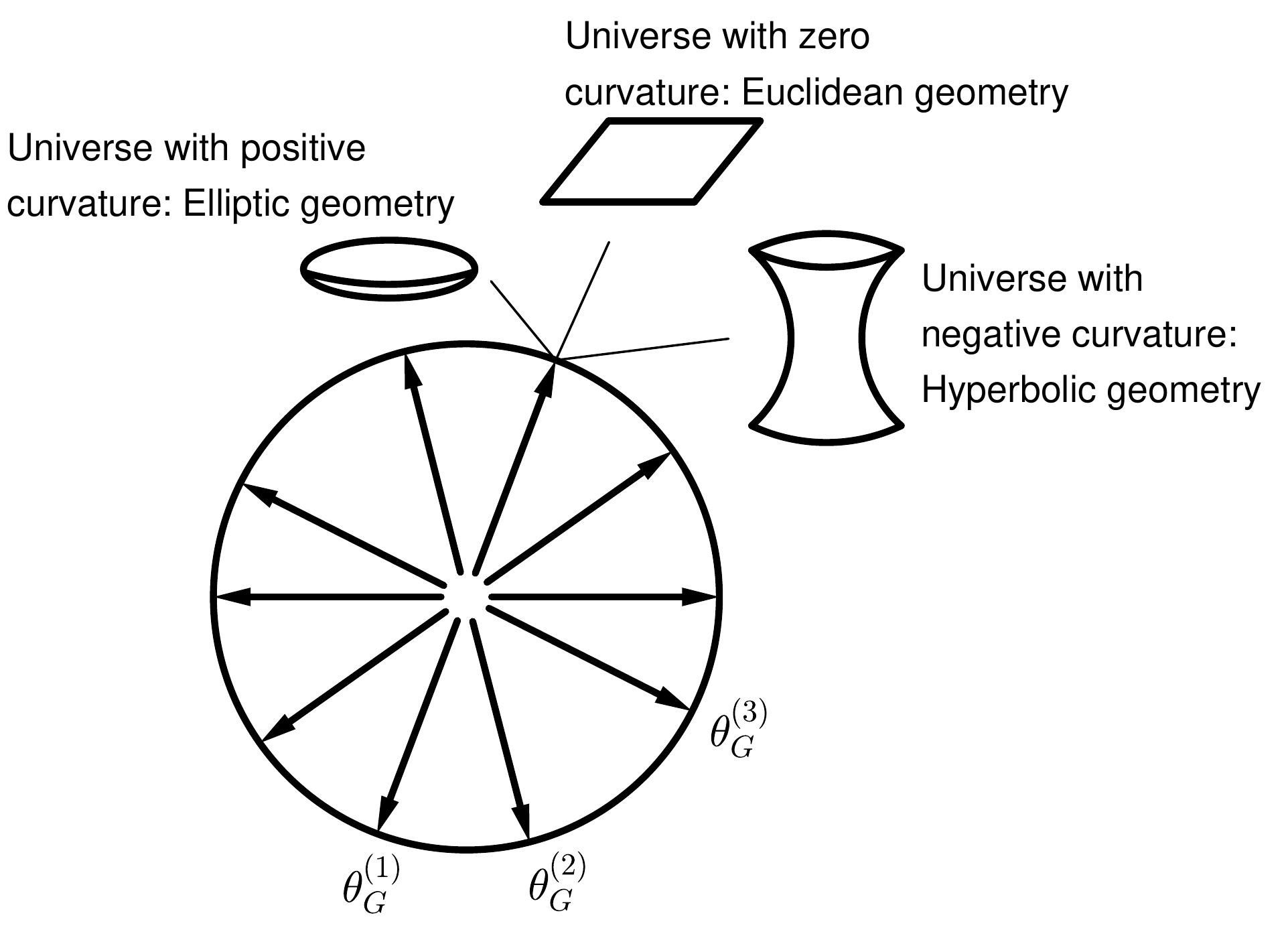}%
\caption{Multiverse's book sheets. We draw for each $\theta_G^{(i)}$ the possible space structures.}
\label{fig2}%
\end{figure}

Indeed, in the multiverse ensemble $\{\Utg\}$, each element, with its own temporal parameter $t_{\theta^{(i)}_G}$ starting from the TLMBB, builds the multiverse's book sheets illustrated in figures~\ref{fig1} and~\ref{fig2}. Each sheet of the book may correspond to negative, zero or positive curvature of the space, although a de Sitter model with elliptic (closed) spatial geometry seems to be the appropriate one for the quantum creation of the observable universe from ``nothing'' (see Sec.~\ref{sec:zero}).
The complete multiverse ensemble $\{\Utg\}$ is larger, in the ordinary sense of geometry, than the observable universe, $\Utgu$, in the same way as a book is more than one of its sheets.

To be more precise, we are following~\cite{Wilczek:2013lra} and ``by universe we mean the domain of physical phenomena which either are, or can reasonably be expected to be, accessible to observation by human beings in the foreseeable future. By multiverse, we mean a larger physical structure of which the universe forms part''.

The main arguments against multiversality~\cite{Tegmark:2003db} are that they are wasteful and weird. The first argument is that multiverse theories are vulnerable to Ockham's razor, since they postulate the existence of other worlds inaccesible to observation by human beings in the foreseeable future. However, as discussed in detail in~\cite{Tegmark:1995dz}, an entire ensemble is often much simpler to accept than one of its members. In fact, if the time coordinate (just as it enters in a Friedmann-Robertson-Walker universe) has an origin, as it is the case in the framework of the mentioned theorems~\cite{Borde:2001nh,Vilenkin:2014yva}, time is better represented as a radial coordinate than otherwise, since in such a case, $t \in [0,\infty)$. But a radial coordinate \emph{has to be} accompanied by angular coordinates, which at this point we generically denote by $\theta_G^{(i)}$. Therefore, the Ockham's razor argument is not telling us that only a sheet of the possible multiverse book, i.e. only a $t_{\theta_G}$, is simpler to accept than the whole book, it would rather be telling us otherwise. Then, extrapolating to our case the Vilenkin picture~\cite{Vilenkin:1982de,Vilenkin:1984wp,Vilenkin:1986cy,Vilenkin:1987kf,Vilenkin:1994rn,Vilenkin:2002ev,Vilenkin:2018dch,Vilenkin:2018oja}, the TLMBB would produce small closed universes which spontaneously nucleate ``out of nothing'' (see Sec.~\ref{sec:zero}): this would be the birth of our multiverse's book sheet.

But we are talking about physics, not mathematics. If the TLMBB had to invest more energy to produce two, three universes, or the entire multiverse, than to produce only our observable universe, then, maybe, there could be energetic arguments that could be used to justify the existence of only our observable universe. Then the Ockham's razor argument would not be saying us that the whole book is easier to accept than only a sheet of the book. But in our multiverse conjecture (see Section~\ref{sec:zero}) it costs the same energy (zero energy) to produce one universe as many, so we maintain that Ockham's razor favors multiversality.

Note that in our scheme we do not accept multiversality forced by the peculiar values of the physical constants of the universe, but by the interpretation of Ockham's razor within our framework. 
Indeed, in a book with a lot of sheets it would not be surprising that in the sheet corresponding to the observable universe the parameters of the standard model would have the values they have, just the right ones needed to allow for complex structures. 
In the present proposal, the multiverse is a natural consequence of the interpretation of time as a radial coordinate, as suggested from the geodesic incompleteness theorems. This interpretation offers then an internal space where multiple time coordinates (corresponding to an ensemble of universes) can be included. Note that this is a motivation for a multiverse much different from those based on the anthropic principle or the eternal inflation paradigm (for a review, see Ref.~\cite{Vilenkin:2006ac}). Note also that this multiverse has not got the problems of other multiverse ideas, such as a zero probability for each universe in a multiverse not labelled by a continuous variable (see page 350 of Ref.~\cite{vilenkin:book}). In short, on the basis that the universe is past incomplete, our multiversality, i.e. a zero energy multiverse ensemble \{$\Utg$\}, is simpler to accept than one of its members.

The second argument, that multiverses are weird, is really a very weird argument: just remember that, for example, liquid Helium II can flow upward; see more examples in Ref.~\cite{Tegmark:2003db}.

The identification of the different sheets of the multiverse's book in Figs.~\ref{fig1} and~\ref{fig2} with an angular $\theta_G$ coordinate is based on the second of the above mentioned conjectures: the physical gravitational topological $\theta_G$-term. 
The motivation for this extension of Einstein gravity will be reviewed in Sec.~\ref{sec:thetag}. Note that such an ingredient allows us to say that quantum chromodynamics (QCD) and QGD have a similar topological structure; note also however that
there is an important distinction between the various $\theta$ vacua of QGD or QCD and the possible numerical values of $\theta$ of a spontaneously broken symmetry, such as the Higgs sector of the electroweak theory. In the latter case, the possible numerical values of $\theta$ label the same theory. In contrast, each value of $\theta$ in QGD or QCD corresponds to a different theory~\cite{Donoghue:1992dd}: the physical gravitational topological $\theta_G$-term specifies the content of the version of QGD used by Nature. 

In section~\ref{sec:zero}, devoted to the zero-energy multiverse, we discuss that, although there are difficulties in obtaining a unique definition of gravitational energy, the use of the Landau-Lifshitz energy momentum pseudo-tensor leads to a negative character of the gravitational energy and to the vanishing of the total energy of closed universes and, with some specifications, to that of spatially flat universes, and even also to that of any FRW universe, including the open universes. In that section we also discuss the zero-energy multiverse created from ``nothing''. By ``nothing'' we mean, according to Vilenkin~\cite{Vilenkin:1982de,Vilenkin:1984wp},
``a state with no classical spacetime and matter'', supported, eventually, in the TLMBB.  How the universe is created by tunneling from ``nothing'' into a perturbative minisuperspace framework of a de Sitter universe is discussed extensively in Refs.~\cite{Vilenkin:1982de,Vilenkin:1984wp,Vilenkin:1986cy,Vilenkin:1987kf,Vilenkin:1994rn,Vilenkin:2002ev,Vilenkin:2018dch,Vilenkin:2018oja}.

In section~\ref{sec:time} we speculate that the TLMBB, where the classical spacetime has disappeared, could be the appropriate world where to build a MQC, and comment on how the ``problem of time''~\cite{Kiefer:1993yg,Kiefer:2009tq,Barbour:1993iv} manifests in our framework. We conclude with section~\ref{sec:concl} by summarizing our results and by considering possible consequences of our proposal. In particular, a footprint of the MQC outside of the TLMBB should be the theory of QG of our observable universe. If this speculation could be implemented, then multiversality would have a predictive power in our observable universe.

\section{The gravitational topological $\theta_G$-term}
\label{sec:thetag}

It is commonly assumed, both that during the Planck epoch the current physical theories do not have predictive value, and that physics is
dominated by the quantum effects of gravity. In this epoch we do not know the physical laws and then we can not label space and time: to all intents and purposes, in this epoch the classical spacetime disappears. In our framework, it disappears at the \cgbb, within which current physics does not work.

Having said this, and in the absence of a full QG theory, we can ask ourselves if any of the proposals in the literature to take into account possible low-energy quantum gravitational effects fits well in our framework. One of these particular effects could be a CP violation induced by the addition to the standard Einstein Hilbert action of a topological $\theta_G$-term
\begin{equation}
S_{\theta_G}\sim \theta_G \int d^4 x \sqrt{-g} \ \epsilon^{abcd}\,R_{abm}^{\ \ \ \ n} R_{cdn}^{\ \ \ \ m},
\label{eq:theta-term}
\end{equation}
where $R_{abm}^{\ \ \ \ n}$ is the curvature tensor, and the quantity $\epsilon^{abcd}\,R_{abm}^{\ \ \ \ n} R_{cdn}^{\ \ \ \ m}$ is (proportional to) what is usually called the \emph{Chern-Pontryagin density}. Such a term can be motivated, in a quite generic way, by the principle of gauge invariance, giving rise to the so-called ``Chern–Simons'' (CS) modified general relativity, where in fact $\theta_G$ can be coordinate-dependent and also evolve dynamically with the inclusion of a new term in the action~\cite{Jackiw:2003pm,Alexander:2009tp}.

There are however important reasons to consider such a term with a constant $\theta_G$ as an ingredient of a ``quantum gravidynamics'' (QGD)~\cite{Deser:1980kc}. In the loop quantum gravity approach, canonical quantization of GR is achieved by analogy to Yang-Mills (YM) theory, which is made manifest when the Einstein-Hilbert action is expressed in terms of Ashtekar ``connection variables'' (essentially, an SU(2) connection $A_a^i$ and its conjugate momenta, the triad, $E^a_i$)~\cite{Ashtekar:1988sw}. The triad must then satisfy the Gauss law, 
\be
\mathcal{D}_a E^a_i=0, 
\label{eq:gauss}
\ee
where $\mathcal{D}$ is the gauge-covariant derivative operator defined by $A_a^i$. 

The constraint Eq.~\eqref{eq:gauss} generates internal gauge transformations in the form of triad rotations, which transform the connection $A\to A'$. As in YM theories, the topology of the quotient of space of connections by the action of the gauge group is nontrivial, and this leads to a 1-parameter ambiguity in the quantization of the theory. In particular, wavefunctions are represented by functionals of $A_a^i$, $\Psi[A]$, which are invariant under ``small'' (continuously connected to the identity) gauge transformations, 
\be
\Psi[A']=\Psi[A],
\ee
while they transform under ``large'' local gauge transformations according to their unitary irreducible representations, which are of the form $e^{in\theta}$, where $n$ is the winding number and $\theta$, $0\leq \theta\leq 2\pi$, is an angular ambiguity parameter. Under a large gauge transformation, 
\be
\Psi[A']=e^{in\theta}\Psi[A], 
\ee
and the quantized theory depends on the value of the $\theta$ parameter. 

It is possible to rescale the wavefunctions to eliminate this $\theta$ dependence, at the cost of changing the expression of the momentum operator, $E_i^a=-i \delta/\delta A_a^i$, with a CP-violating $\theta$-dependent term. 
This redefinition of the momentum can be in fact understood at the level of the action (on solutions to the field equations~\cite{Ashtekar:1988sw,Calcagni:2017sdq}) as arising from the addition of the topological term Eq.~\eqref{eq:theta-term} with a constant (coordinate independent) value of $\theta_G$~\cite{Alexander:2009tp,Ashtekar:1988sw}. Topological considerations of QGD, therefore, lead to a 1-parameter family of quantum theories (labelled by $\theta_G$) with P and CP violation in every sector except for $\theta_G=0,\pi$~\cite{Ashtekar:1988sw}. In summary, loop quantum gravity gives base to the term Eq.~(1)~\cite{Rovelli:2004tv}.

The parity violating gravitational $\theta_G$-term Eq.~\eqref{eq:theta-term} has been studied in connection with black-hole physics, where it has been show that, although this term does not contribute to the classical equations of motion because it is a total derivative, it affects the transport properties of the horizon, suggesting that the $\theta_G$-term may play an important role in a sensible theory of QG~\cite{Fischler:2016jaq,*Fischler:2015kro}. Some implications of this topological term in different extensions of gravity have also been considered recently~\cite{Obregon:2012zz,Chagoya:2016zhy}. At this point we want to remark that such an ingredient naturally fits in our model, since the arbitrariness in the $\theta_G$ parameter provides us with an angular coordinate which distinguishes a specific universe, as shown in Fig.~\ref{fig1}. It is therefore a particular realization of the ensemble of universes that arises in our framework, when the geodesic incompletitude theorems forces one to consider time as a radial coordinate. In the following section we turn our attention to the multiverse idea.

\section{Negative character of the gravitational energy and the zero-energy multiverse of quantum cosmology}
\label{sec:zero}

Our zero-energy multiverse conjecture has two parts. The first one is that, although there are difficulties in obtaining a unique definition of gravitational energy, the use of the Landau-Lifshitz energy momentum pseudo-tensor~\cite{landau} leads to a negative character of the gravitational energy and to the vanishing of the total energy of closed universes and, with some specifications~\cite{Johri:1995gh}, to that of spatially flat universes and even also to that of any FRW universe, including the open models~\cite{Cooperstock95}. It is worthwhile to mention that the vanishing of the total energy of the closed universe as calculated in Ref.~\cite{Johri:1995gh}, by using the Landau-Lifshitz pseudo-tensor, corroborates Rosen's results~\cite{Rosen:1994vj} based on the Einstein pseudo-tensor and confirms also the views of Guth~\cite{Guth:1993sz}, Hawking~\cite{Hawking:2005yn} and Cooperstock~\cite{Cooperstock:1994zu}.
It is also interesting to note that, without resorting to mathematics, Guth explains in the appendix quoted in Ref.~\cite{Guth:1997wk}, ``how the properties of gravity can be used to show that the energy of a gravitational field is unambiguously negative''.

This is fine, but during the Planck epoch the current physical theories (on which these previous results are based, Refs.~\cite{landau,Johri:1995gh,Cooperstock95}) do not have predictive power because at that moment physics is dominated by quantum effects of gravity, which are not taken into account by these theories. Nevertheless, there are earlier proposals that the universe arose as a quantum fluctuation of the vacuum, which also implies zero value for the energy of the universe~\cite{Hawking:2005yn,Albrow:1973wp,Tryon:1973xi,Brout:1977ix,Vilenkin:1982de,Vilenkin:1984wp,Vilenkin:1986cy,Vilenkin:1987kf,Vilenkin:1994rn,Vilenkin:2002ev,Vilenkin:2018dch,Vilenkin:2018oja,Pollock:1985vk}. To our knowledge, the idea that our universe might be a vacuum fluctuation was originally suggested, first in a vague way by Albrow~\cite{Albrow:1973wp}, and then more explicitly by Tryon~\cite{Tryon:1973xi}. 
It is in fact more accurate to say that the universe is created from ``nothing'', where by ``nothing'' we mean, according to Vilenkin, ``a state with no classical spacetime and matter''~\cite{Vilenkin:1982de,Vilenkin:1984wp,Vilenkin:1986cy,Vilenkin:1987kf,Vilenkin:1994rn,Vilenkin:2002ev,Vilenkin:2018dch,Vilenkin:2018oja}. We recommend chapters 16 and 17 of Ref.~\cite{Vilenkin:2006ac} for a very clear introduction to how the universe is created by quantum tunneling from ``nothing'' into a de Sitter space. The ``tunneling'' wave function approach to quantum cosmology, using a simple model of a closed Friedmann-Robertson-Walker universe, and comments on the alternative proposals for the wave function, are discussed in Refs.~\cite{Vilenkin:2002ev,Vilenkin:2018oja}, while a more complete analysis of the tunneling wave function of the universe is given in the recent Ref.~\cite{Vilenkin:2018dch}. Note that this \emph{creation} of the universe is unavoidable, since, as emphasized in note 8 of chapter 16 of Vilenkin's book~\cite{Vilenkin:2006ac}, ``spacetime itself is past-incomplete, and therefore does not provide a satisfactory model of a universe without a beginning.'' In our framework, which is supported on a \cgbb, spacetime is indeed past-incomplete, in agreement both with the comment in Vilenkin's book and the aforementioned observation by Hawking~\cite{Hawking:2005yn}.

The second part is to make an obvious use of these results in our zero-energy multiverse. Our conjecture is essentially ``the conjecture of the multiverse of quantum cosmology'', level 3 of the multiverse in the terminology of Perlov and Vilenkin~\cite{vilenkin:book}: ``Level 3: multiple disconnected spacetimes produced by quantum tunneling from nothing''. Note also that in the quantum tunneling from ``nothing'' proposal, a universe can emerge with any of a variety of values for the vacuum energy~\cite{vilenkin:book}: ``We shall refer to this ensemble of universes as the multiverse of quantum cosmology''.
In our case, we assume that the complete multiverse ensemble, \{$\Utg$\}, arising out from the \cgbb, is produced by quantum tunneling from nothing. In our multiverse conjecture it costs the same energy (zero energy) to produce one universe as many, so that we maintain that Ockham's razor favors multiversality (see section~\ref{sec:intro}).

In Ref.~\cite{vilenkin:book}, it is also recalled that ``the level 3 multiverse of quantum cosmology is regarded as an intriguing possibility, but progress in this area should await the development of the theory of quantum gravity''. But, what if things were the other way round and an identification of the framework in which to build a QC were needed as a previous step to develop QG? As Wiltshire says~\cite{Wiltshire:1995vk}, referring to QC and QG: ``Quantum cosmology is perhaps most properly viewed as one attempt among many to grapple with the question of finding a quantum theory of gravity''. If this were the case, the first thing to do should be to imagine the framework for QC. A past-incomplete universe~\cite{Borde:2001nh,Vilenkin:2014yva} invites to consider our proposal as a reasonable framework.

Following Wilcek's definition~\cite{Wilczek:2013lra}, the multiverse is a larger physical structure of which our universe forms part. However, in our framework we have two different regions, one outside the \cgbb, determined by all the elements of our 
multiverse book, \{$\Utg$\}, and another one corresponding to the unknown nature of the \cgbb, in which all members of the multiverse ensemble, together with their respective classical spacetimes, disappear (see the next section). The physical or mathematical underlying construction for the \cgbb\ which would be needed to build the MQC (the analogous of Hilbert spaces in the case of quantum mechanics, for example) is then an open question that, as we argue in Sec.~\ref{sec:concl}, could be related to the formulation of QG.

\section{Additional comments and the problem of time}
\label{sec:time}

Usually we base quantum cosmology on a theory of quantum gravity, mostly quantum geometrodynamics or loop quantum gravity (LQG). These approaches to the quantization of gravity are based on the Wheeler-DeWitt (WD) equation. For some authors (see Ref.~\cite{Isham:1992ms}), the WD equation (elegant though it be) suffers from the problem of time and so it may be the ``wrong way of formulating a QG''. For others, the WD equation is more fundamental than the Schrödinger equation. This point of view is closely related to the idea that time is something irrelevant in QG~\cite{Kiefer:1999es}: ``In classical canonical gravity, a spacetime can be represented as a \emph{trajectory} in configuration space ---the space of all three-metrics (...) Since no trajectories exist anymore in quantum theory, no spacetime exists at the most fundamental, and therefore, also no time coordinates to parameterize any trajectory.'' In two words, time is absent because classical spacetime vanishes upon quantization in the same way as a particle trajectory in quantum mechanics, see also~\cite{Rovelli:2004tv}. These two options have been discussed recently in Ref.~\cite{Shestakova:2018crq}.

Where does then time come from in quantum geometrodynamics? Usually, it is an emergent quantity for which one needs two conditions: the validity of a Born-Oppenheimer type of approximation, and decoherence~\cite{Kiefer:1993fg}. Decoherence is discussed, for example, in sections 4.1 and 4.2 of Ref.~\cite{Kiefer:1993fg}, within the framework of the WKB expansion~\cite{Kiefer:1990pt}. In these sections it is demonstrated that the use of a single WKB state can be justified in a natural, physical way, which allows one to promote the WKB time to a physical time (see also~\cite{Kiefer:1993yg,Kiefer:2009tq,Barbour:1993iv}).

In the previous discussion, quantum cosmology is based on a theory of quantum gravity. Alternatively, in this work we have asked ourselves:  what if things were the other way round, and an identification of the framework in which to build a QC were needed as a previous step to develop QG? In our framework, all members of the multiverse ensemble, together with their respective classical spacetimes, disappear at the TLMBB. The TLMBB is then the realm of the MQC, and the semiclassical disconnected universes should be in fact connected through quantum entanglement at this region.

Now the concern is, where do the temporal parameters of our multiverse ensemble come from? Making it clear that we do not have the answer, based on what we have just said, one of the first steps to consider would be the construction of an appropriate representation space for the dynamical variables which were able to take into account the apparition of the $\theta_G^{(i)}$. How this would affect the subsequent WKB program and would characterise the emergence of our temporal parameters (ie, the emergence of the sheets or semiclassical domains of our multiverse) that occur after the ``initial'' TLMBB is an open question.

In the next paragraph we will refer to a correspondence principle that any physical or mathematical structure underlying the TLMBB should incorporate.

\section{Conclusions}
\label{sec:concl}

Our proposal of a framework for MQC has been based on three cosmological conjectures that give the main features of this framework: the geodesic incompleteness theorems led us to consider a \cgbb, with a radial interpretation of the time coordinate, which implies an ensemble of universes; this ensemble has a natural realization in the one-parameter family of inequivalent vacua in QGD; finally, the existence of this set of universes, which are labelled by the topological $\theta_G$-term, is made consistent in a zero-energy multiverse. We will now finish by commenting on some features and possible consequences of this proposal.

First, it is interesting to note that the angular coordinate that arises naturally in our framework can be related to a theoretical angular ambiguity in the formulation of loop quantum gravity~\cite{Rovelli:2004tv}, producing a CP-violating term that may be relevant to QG, as the black-hole analysis of Ref.~\cite{Fischler:2015kro} suggests.

Secondly, we would like to remark that the approach of Vilenkin~\cite{Vilenkin:1982de,Vilenkin:1984wp,Vilenkin:1986cy,Vilenkin:1987kf,Vilenkin:1994rn,Vilenkin:2002ev,Vilenkin:2018dch,Vilenkin:2018oja}, based on the picture that small closed universes spontaneously nucleate out of nothing, fits well in our framework~\cite{Vilenkin:1994rn}: ``All universes in this metauniverse are disconnected from one another, and generally have different values for some of the constants. This variation may be due to different compacting schemes, etc (...) After nucleation, the universes enter a state of inflationary expansion.'' Note that the generalization of Vilenkin's tunneling proposal to our multiverse framework is straightforward. Indeed, it is enough to incorporate into our scheme the creation of the multiverse in entangled pairs (see below) and take into account that the classically forbidden region of Vilenking's tunneling proposal occurs within our TLMBB phase (see what follows).

Vilenkin's tunneling is basically different from tunneling in ordinary quantum mechanics, where it takes place from one classical allowed region to another classically allowed region. The transition from ``nothing'' to Vilenkin's universe, however, is supposed to take place from a classically forbidden (Euclidean) region to a classically allowed (Lorentzian) region (see for example~\cite{Vilenkin:1984wp,Vilenkin:1987kf,Gott:1997pm}), so that the conservation of current is obviously violated. Section IV of Ref.~\cite{Gott:1997pm} discusses this point in detail. How would this violation fit in the proposal presented here? In relation to this question, one thing to keep in mind is that from the point of view of a MQC, the classically disconnected universes could be in fact connected through quantum entanglement, as we have discussed in section~\ref{sec:time}. In our case, the decoherence required to arrive at the semiclassical domains after the ``initial'' TLMBB may not be complete and it may indeed have some (potentially observable) signatures of entanglement of our universe with other semiclassical domains. Although interesting, in the absence of knowing how to make a realistic calculation, this, at the moment, is a very speculative issue. Nevertheless, it could be mentioned that in the context of minisuperspace models~\cite{Kiefer:2007ria,Kuchar:1991qf}, it can be argued that if the universes were created in entangled pairs with opposite values of the momenta conjugated to the configuration variables of the minisuperpace, then the aforementioned violation of the current would not occur~\cite{Robles-Perez:2018vcq}.

The idea of creation of a pair of universes has also been recently explored in~\cite{Boyle:2018tzc,Boyle:2018rgh}, in connection with CPT symmetry at the level of the universe. In their CPT-symmetric model, the authors consider a spacetime covered by a coordinate $\tau$ which runs from $-\infty$ to $+\infty$, and the CPT symmetry is manifest by the $\tau\to-\tau$ isometry. As the authors point out, this spacetime can be interpreted as two parallel universes (a universe and anti-universe pair) which emerge from the origin. We want to remark that this interpretation, in which time coodinate is positive in both universes, can be immediately associated to a one-dimensional slice of our Fig.~\ref{fig1}. It is interesting that it is possible to link certain experimental observations to this idea~\cite{Anchordoqui:2018ucj,Anchordoqui:2018qom}. 

It is also interesting to note that the regularization procedure of the big bang singularity described in Ref.~\cite{Klinkhamer:2019dzc} suggests the existence of a pre-big-bang phase, in our case, of the TLMBB phase. In fact, the assumption behind such regularization procedure is that quantum mechanical effects tempers the divergences associated to the big bang, making this phase to play the role of a ``quantum bridge'' between, in our case, the different elements of our multiverse.

Finally, general relativity and cosmology consider the large scale structure of spacetime. To quantize GR, one has to quantize spacetime itself, rather than the fields that live in that spacetime. On the other hand, QC is viewed as an attempt among many to face with the question of finding a theory of QG~\cite{Wiltshire:1995vk}. Therefore, a framework for (multiverse) quantum cosmology, as the one we are proposing here, based on three well-studied cosmological conjectures, could help us in the search for QG. In Section~\ref{sec:zero}, we have paid attention to this possibility when we considered that the identification of the appropriate framework to build a QC could be a necessary step before developing a theory of QG. Now, any physical or mathematical structure underlying the \cgbb\ should incorporate a correspondence principle (quite speculatively, and as an example, an attempt could proceed along the line of Ref.~\cite{Singh:2017ved}) such that the MQC theory should lead, at low energies, to the extension of Einstein's gravity by a physical gravitational topological $\theta_G$-term~\cite{Deser:1980kc,Ashtekar:1988sw} outside the \cgbb. In this case, as our proposal would have turned out to be the detonator of this property, we would expect that multiversality would provide us with some pieces of the still unknown theory of QG.

\section*{Acknowledgments}
We would like to thank Filiberto Ares, Carlos Bouthelier, José Luis Cortés, Jorge Alberto Jover, and Amilcar R. de Queiroz for discussions. We would also like to thank Latham Boyle, Frans R. Klinkhamer, and Salvador Robles-Pérez, for very fruitful correspondence. A special mention deserves our thanks to Claus Kiefer, for a detailed reading of our work and for suggesting us to add a new section commenting on how the ``problem of time'' manifests in our framework.

\end{document}